\documentclass[sigconf]{acmart}
\AtBeginDocument{%
  }

\acmConference[SEIP-ICSE 26]{}{April,
  2026}{BRAZIL}
\acmISBN{978-1-4503-XXXX-X/2018/06}

\usepackage{multirow}
\usepackage{graphicx}
\usepackage{pgfplots}
\pgfplotsset{compat=1.18}

\begin{document}

\title{Deriving and Validating Requirements Engineering Principles for Large-Scale Agile Development: An Industrial Longitudinal Study}

\author{Hina Saeeda}
\authornotemark[1]

\affiliation{%
  \institution{Chalmers University of Technology}
  \city{Gothenburg}
  \state{}
  \country{Sweden}
}
\email{hinasa@chalmers.se}

\author{Mijin Kim}
\affiliation{%
  \institution{University of Gothenburg}
  \city{Gothenburg}
  \country{Sweden}}
\email{gusmijki@student.gu.se}

\author{Eric Knauss}
\affiliation{%
  \institution{University of Gothenburg}
  \city{Gothenburg}
  \state{}
  \country{Sweden}}
   \email{eric.knauss@cse.gu.se}

\author{Jesper Thyssen}
\affiliation{%
 \institution{Grundfos Holding A/S}
 \city{Bjerringbro}
 \state{}
 \country{Denmark}}
 \email{jthyssen@grundfos.com}

 \author{Jesper Ørting}
\affiliation{%
 \institution{Grundfos Holding A/S}
 \city{Bjerringbro}
 \state{}
 \country{Denmark}}
 \email{joerting@icloud.com}

\author{Jesper Lysemose Korsgaard}
\affiliation{%
 \institution{Grundfos Holding A/S}
 \city{Bjerringbro}
 \state{}
 \country{Denmark}}
  \email{ jlysemose@grundfos.com}

\author{Niels Jørgen Strøm }
\affiliation{%
 \institution{Grundfos Holding A/S}
 \city{Bjerringbro}
 \state{}
 \country{Denmark}}
  \email{njstroem@grundfos.com.}

\renewcommand{\shortauthors}{Saeeda et al.}

\begin{abstract}

In large-scale agile systems development, the lack of a unified requirements engineering (RE) process is a major challenge, exacerbated by the absence of high-level guiding principles for effective requirements management.  To address this challenge, we conducted a five-year longitudinal case study with \href{https://www.grundfos.com/}{\underline{Grundfos AB}}, in collaboration with the \href{https://www.software-center.se/}{\underline{Software Centre}} in Sweden. RE principles were first derived through qualitative data collection spanning more than $25$ sprints, approximately $320$ weekly synchronization meetings, and seven cross-company/company-specific workshops between $2019$–$2024$. These activities engaged practitioners from diverse roles, representing several hundred developers across domains. In late $2024$, five in-depth focus groups with senior leaders at Grundfos provided retrospective validation of the principles and assessed their strategic impact. We aim to (1) empirically examine RE principles in large-scale agile system development, (2) explore their benefits in practice within the case company, and (3) identify a set of transferable RE principles for large-scale contexts. Using thematic analysis, six key RE principles—\textit{architectural context, stakeholder-driven validation and alignment, requirements evolution with lightweight documentation, delegated requirements management, organisational roles and responsibilities, and shared understanding of requirements} are derived. The study was further validated through cross-company expert evaluation with three additional multinational organisations (Bosch, Ericsson, and Volvo Cars), which are directly responsible for large-scale requirements management. Together, these efforts provide a scalable and adaptable foundation for improving requirements practices in large-scale agile organisations.

\end{abstract}

\begin{CCSXML}
<ccs2012>
 <concept>
  <concept_id>00000000.0000000.0000000</concept_id>
  <concept_desc>Do Not Use This Code, Generate the Correct Terms for Your Paper</concept_desc>
  <concept_significance>500</concept_significance>
 </concept>
 <concept>
  <concept_id>00000000.00000000.00000000</concept_id>
  <concept_desc>Do Not Use This Code, Generate the Correct Terms for Your Paper</concept_desc>
  <concept_significance>300</concept_significance>
 </concept>
 <concept>
  <concept_id>00000000.00000000.00000000</concept_id>
  <concept_desc>Do Not Use This Code, Generate the Correct Terms for Your Paper</concept_desc>
  <concept_significance>100</concept_significance>
 </concept>
 <concept>
  <concept_id>00000000.00000000.00000000</concept_id>
  <concept_desc>Do Not Use This Code, Generate the Correct Terms for Your Paper</concept_desc>
  <concept_significance>100</concept_significance>
 </concept>
</ccs2012>
\end{CCSXML}

\ccsdesc[500]{Software and its engineering~Requirements analysis}
\ccsdesc[500]{Software and its engineering~Agile software development}
\ccsdesc[500]{Software and its engineering~Software development process management}

\keywords {Requirements engineering, Requirements principle, Large-scale agile, Systems engineering, Large-scale systems development, Process improvement, Requirements engineering practices.} 


\maketitle

\section{Introduction}

Agile methods have enhanced large-scale projects by promoting adaptability, continuous delivery, and collaboration \cite{saeeda2023challenges}, yet their implementation in such contexts is more complex than in traditional approaches \cite{r4dikert2016challenges}. Challenges in large-scale agile development include communication, coordination, inconsistent agile practices, and issues in requirements engineering (RE) and quality assurance \cite{r4dikert2016challenges, orosz2023scaling}. RE in large-scale systems must address complexity, stakeholder alignment, and evolving requirements \cite{saeeda2023challenges}. High-level RE guidelines are crucial for managing these challenges by structuring requirements, supporting inter-team collaboration, and fostering a shared understanding across diverse organizational roles \cite{Hood2008}.

Although large-scale agile software development has been extensively studied \cite{saeeda2023challenges}, there is limited focus on large-scale agile systems development from a requirements engineering (RE) perspective. Kasauli et al. \cite{kasauli2017requirements} identified RE challenges and potential solutions in such contexts, but a gap remains for high-level RE principles. These principles, unlike rigid processes, should function as adaptable guiding policies \cite{ozdenizci2021business}, enabling organizations to tailor RE practices to varied settings, which is crucial in complex, multi-team agile environments. The International Requirements Engineering Board (IREB) defines RE principles as overarching guidelines for activities such as planning and analyzing requirements \cite{IREB2024}. Yet, their general nature lacks specificity for large-scale agile systems and cross-disciplinary alignment.

Large-scale system development introduces complexities that exceed those in traditional or small-scale agile settings. These include the integration of diverse hardware and software components, geographically distributed teams, and varied customer needs \cite{Dingsoyr2014,Eklund2014}. Cross-disciplinary coordination is particularly challenging, especially when balancing system-wide consistency with team-level autonomy \cite{saeeda2023challenges}. Additional concerns such as evolving system architectures, long product lifecycles, regulatory compliance, and the tension between innovation and stability further complicate development \cite{r9edison2021comparing}. These factors necessitate RE practices that support architectural traceability, distributed responsibilities, and ongoing requirement refinement \cite{saeeda2023challenges,Dingsoyr2014}. Effectively applying RE principles can improve cross-team collaboration and help ensure that complex customer needs are consistently addressed at the system level \cite{r9edison2021comparing}.

This study seeks to explore guiding principles for large-scale agile system development, which are currently lacking. 
To achieve the goals, this study addresses the following research questions:

\begin{itemize}
   \item \textbf{RQ1:} What could a suitable set of overarching \textit{RE principles} look like in a large-scale systems development company?

    \item \textbf{RQ2:} What are the perceived benefits of defining or aligning on  \textit{RE principles} in a large-scale systems engineering organization?
    
    \item \textbf{RQ3:} What \textit{RE principles} for large-scale agile systems development can be recommended based on literature and practical experience?

\end{itemize}

Unlike prior work (e.g., \cite{kasauli2017requirements, Dingsoyr2014, r9edison2021comparing}) that discusses individual aspects of RE, this five-year longitudinal study contributes a synthesised and empirically grounded set of principles tailored for large-scale agile system development. Following a methodological triangulation approach, which combines empirical data, literature insights, and industrial validation, the study provides practical and actionable RE guidance for complex, large-scale agile environments.

\section{Related Work}

\textbf{Agile way of working and RE}: RE and agile are often seen as incompatible due to differences in managing requirements and responsibilities\cite{kasauli2017requirements}. Traditional RE uses experts in a dedicated phase with extensive documentation, while agile continuously manages requirements with minimal documentation through customer-team communication\cite{Inayat2014}. Despite these differences, studies show both challenges and benefits in combining them\cite{saeeda2023challenges}. Agile helps address over-scoping and improves communication, fostering shared understanding\cite{ Dingsoyr2014}. However, maintaining customer communication and minimal documentation can be difficult, especially for organizations needing detailed records\cite{Inayat2014}. Balancing agile flexibility with RE stability is essential for success\cite{saeeda2023challenges}.

\textbf{Large-scale agile development and RE}: The literature defines large-scale agile by i) the organization's size, ii) team size, and iii) the number of teams involved. Dingsøyr et  al.\cite[p.~275]{Dingsoyr2014} describe large scale as \textit{2-9 collaborating teams, with over ten being very large}. Dikert et al. \cite[p.~88]{r4dikert2016challenges} define large-scale as \textit{involving 50 or more people or at least six teams}.

Large-scale agile development faces challenges like a lack of shared vision, limited guidance, and team communication issues, which also affect RE\cite{Inayat2014}. A study by Kasauli et al. \cite{kasauli2017requirements} identified 24 RE challenges in large-scale agile systems, including shared customer value, integration with existing structures, and consistent requirements usage across teams.

\textbf{Established Frameworks in RE for Large Scale Agile:}

Several established frameworks provide important foundations for requirements engineering (RE) in agile and large-scale development contexts. These frameworks served as a baseline for this study, informing both our analysis and validation of RE principles.  

\textbf{(I) IREB Principles.}  
The internationally recognised RE handbook by IREB \cite{IREB2024} outlines nine fundamental principles of requirements engineering. These include: (1) Value orientation – using requirements to create value by reducing costs and increasing benefits; (2) Stakeholder – considering and harmonizing different stakeholders’ needs and wants; (3) Shared understanding – ensuring a common understanding among all parties involved; (4) Context – understanding the system’s environment and defining relevant boundaries; (5) Problem, requirement, solution – using requirements to capture problems and guide solution development; (6) Validation – continuously verifying that requirements meet stakeholder needs; (7) Evolution – recognizing that requirements change over time; (8) Innovation – encouraging improvements beyond stated needs; and (9) Systematic and disciplined work – conducting RE activities in an organized manner while allowing for agility and flexibility. While comprehensive, they are intended as general-purpose guidelines and are not specifically tailored to the challenges of large-scale agile development.  

\textbf{(II) Kasauli et al.’s Taxonomy of RE Challenges.}  
Kasauli et al.\cite{kasauli2017requirements}  developed a taxonomy of $24$ RE challenges in large-scale agile system development, grouped into six overarching themes: \textit{process, communication, roles and responsibilities, artefacts, system-level alignment, and tool support}. This taxonomy highlights the complexity of RE in large-scale settings, especially the tension between local team autonomy and global system coordination. It provides an empirically grounded reference point for analysing challenges that emerge in industrial agile environments.  

\textbf{(III) Inayat et al.’s Agile RE Practices.}  
Inayat et al.~\cite{Inayat2014} conducted a systematic review of agile RE practices, identifying recurring approaches such as lightweight documentation, user stories, continuous stakeholder involvement, prioritisation, and iterative validation. These practices provide insights into how agile teams handle RE in practice but are mainly synthesised from project-level studies, with less emphasis on scaling to complex, multidisciplinary environments.  

\textbf{Positioning RE principles.}  
Our study builds on these frameworks in two ways. First, IREB provided a normative baseline for mapping and interpreting initial findings. Second, Kasauli et al.’s taxonomy and Inayat et al.’s agile practices offered complementary perspectives specifically attuned to the challenges of agile RE at scale. By situating our derived principles in relation to these frameworks, we ensure that they are not only consistent with established knowledge but also extend it toward addressing the realities of large-scale, multidisciplinary system development.

\section{Research Methodology}

Following the case study research approach proposed by Runeson et al.~\cite{Runeson2009}, we conducted a longitudinal industrial case study with \href{https://www.grundfos.com/}{\underline{Grundfos AB}~}, covering approximately five years of organizational transition. Our engagement spans the company’s shift from stage-gate practices toward agile methods beginning in $2019$, the subsequent scaling of agile to more than $100$ teams by $2022$, and the ongoing validation of requirements engineering (RE) principles to address the challenges of large-scale, multidisciplinary system development. 

The study was conducted as part of the \href{https://www.software-center.se/}{\textbf{\underline{Software Center}}}’s \href{https://www.software-center.se/intranet/projects/re-for-large-scale-agile-system-development-27/}{\textbf{\underline{Project~27:-}~}} \textit{Requirements Engineering for Large-Scale Agile System Development}, with initial project activities documented in the \href{https://www.software-center.se/wp-content/uploads/2020/05/SC-AR-2019-FINAL.pdf}{\textbf{\underline{Software Center Annual Report 2019}~}}.

\subsection{Company Context}

 Grundfos AB is a mature, leading global pump manufacturer headquartered in Bjerringbro, Denmark. Founded in 1945 by Poul Due Jensen and owned by the Grundfos Foundation. As of 2022, Grundfos was represented by more than 100 companies in over 60 countries, with products distributed even more widely through local partners. The company employs approximately 19,000 people globally and produces around 15 million pumps annually.
In 2019, Grundfos embarked on a large-scale agile transformation, moving away from a stage-gate model toward practices such as Scrum, Kanban, and, in some programs, SAFe and Continuous Delivery.

Grundfos employs more than 1000 developers in Denmark, organised into approximately 100 agile teams working across hardware, firmware, cloud, and system integration domains.

While the agile shift increased flexibility and accelerated delivery, it also exposed significant challenges in managing requirements across hundreds of developers and dozens of teams. Key issues included the misalignment of requirements between hardware and software teams, inconsistent documentation practices, unclear ownership of system-level requirements, and difficulty maintaining traceability across long product lifecycles and regulatory contexts.

These challenges underscore the need for a set of consistent, organisation-wide RE principles to provide lightweight, adaptable guidance capabilities that are not sufficiently addressed by agile methods alone. Importantly, the solution was not to introduce yet another process. Processes are inherently context-specific and lose their utility when abstracted; tailoring is required in every new setting. By contrast, principles and values can be generic, serving as high-level guiding policies that inform processes without constraining them. Whereas a process prescribes how to act in a fixed context, principles shape the outcomes of processes, ensuring alignment, knowledge retention, and effective coordination while preserving the flexibility central to agile development.

\subsection{Project Context}
Project~27, initiated in 2019 within the Software Centre, addressed a key bottleneck in continuous software engineering: managing requirements and knowledge across agile teams and system levels. For Grundfos AB, the project supported improvements in development speed, product quality, and strategic alignment by providing a structured path toward organisation-wide RE principles. Over more than five years and 25+ sprints, the collaboration included weekly synchronization meetings, sprint reviews, retrospectives, and milestone workshops, ensuring both rigor and industrial relevance. Importantly, Project 27 involved not only Grundfos but also Bosch, Ericsson, and  Volvo Cars, enabling validation across multiple companies and strengthening the generalisability and transferability of the resulting RE principles.

\subsection{Multi-Phase Research Approach}
\textbf{Conducted qualitative data collection }continuously over five years (2019–2024) through more than 25 sprints, approximately 320 weekly synchronisation meetings, and seven cross-company/company-specific workshops with  Software Centre partners and  Grundfos leaders. These activities engaged practitioners from diverse roles (developers, product owners, architects, and managers), representing hundreds of developers across domains. In late 2024, a final set of five in-depth focused groups with senior leaders at Grundfos was conducted to retrospectively validate the formulated RE principles and assess their strategic impact.

\textbf{Literature synthesis}:- Carried out iteratively in parallel with the project’s empirical activities, involving more than ten researchers (senior scholars and early-career researchers). This process systematically connected empirical findings to existing research, and it was continuously updated as new results from Project~27 and related Software Centre studies were published. (See the literature published by Project 27
\href{https://www.software-center.se/intranet/projects/re-for-large-scale-agile-system-development-27/swc27-publications-presentations/}{\textbf{\underline{ Publications and Presentations}}}).

As new literature was published based on the project increments, the existing state-of-the-art literature was utilised for theory building \cite{Inayat2014,kasauli2017requirements}.

\textbf{Cross-company expert validation}:- Besides four senior leaders at Grundfos, we involved selected experts of three additional large system development organisations that were directly responsible
for processes, methods, and tools for requirements management in large-scale
system development in their companies. As the main validation was performed by Grundfoss experts, these additional three experts were engaged in discussion and reasoning for the validation of the choices. Their participation ensured that the derived RE principles were not only validated in the Grundfos case but also benchmarked against practices in multiple multinational organisations, thereby strengthening their generalisability and transferability. Together, these efforts supported the derivation and evaluation of RE principles tailored for large-scale agile system development. 

\subsection{Data collection
}

This study employed various data collection methods, as shown in Fig.~\ref {fig:data_analysis}. 

\textbf{Focus group interview }was conducted to answer RQ1 (Example of RE principle) and RQ2 (Benefits of RE principles). Researchers served as interviewers and moderators, while a leading contact at Grundfos AB recruited five information-rich participants (directly leading and managing 500 software engineers involved with the solution part of the RE-related challenges) with relevant knowledge and experience through purposeful sampling \cite{Runeson2009}. The two-hour, semi-structured session was conducted via online video call with four current employees and one former employee, all of whom are founders or active users of the RE principles. The term ``RM" was used for RE, as is standard at Grundfos AB.
The \noindent \href{ https://doi.org/10.7910/DVN/RYHXJY}{\textcolor{blue}{\textit{\underline{Interview guide}}}} was developed collaboratively by researchers, was pilot-tested, reviewed by a senior researcher, and shared with participants a week in advance. It contained $12$ questions to explore RE principles and company context. The focus group followed this guide flexibly to allow open discussion.
\begin{table}[t]
  \centering
  \caption{Workshop participants}
  \begin{tabular}{lll} 
    \toprule
    \textbf{Participant} & \textbf{Domain} & \textbf{Role} \\
    \midrule
    P1 & Telecommunication & Systems Engineer \\
    P2 & Telecommunication & RM Architect \\
    P3 & Automotive & Requirement Engineer \\
    P4$^1$ & Systems manufacturing & Systems Engineer \\
    \bottomrule
  \end{tabular}
  \label{tab:participants}
      \\[0.1cm] 
  \footnotesize{$^1$ P4 is one of the founding members of the \textit{RE principles } at Grundfos AB and has an Agile Coach role too. }
\end{table}

\textbf{Literature findings}  To strengthen Grundfos AB’s RE principles, a semi-systematic literature review was conducted across IEEE Xplore, Scopus, and ACM Digital Library, focusing on RE in large-scale and agile contexts. While most sources provided practical guidance, the IREB handbook\cite{IREB2024} stood out for its overarching RE principles. These were compared with Grundfos AB’s principles and presented during a cross-company workshop, informing subsequent interviews. Insights from prior Software Center research\cite{kasauli2017requirements, WohlrabRebekka2019Boat,treqs} (Project 27) further supported the formulation and empirical refinement of candidate principles.

\textbf{Workshop} In the final phase of the study, a 3.5-hour cross-company validation workshop was held at the \href{https://www.software-center.se/}{Software Center}. Four senior leaders from \href{https://www.grundfos.com/}{Grundfos AB}, overseeing over 500 engineers, represented the organisation's practices.  Participants rated each RE principle on a five-point Likert scale and engaged in structured discussions. Quantitative analysis (mean, mode, standard deviation) was used to rank principles and assess agreement levels, complemented by qualitative reasoning that explored contextual nuances. In addition to Grundfos, representatives from Bosch, Ericsson, and Volvo Cars contributed by comparing the principles with their own contexts. Despite differences across automotive, manufacturing, and telecommunications domains, many principles were found relevant, confirming their generalisability beyond a single company (see Tables~\ref{tab:workshop} and~\ref{tab:participants}).

\subsection{Data Analysis} 
\begin{figure*}[ht]
\centering
\includegraphics[width=\textwidth]{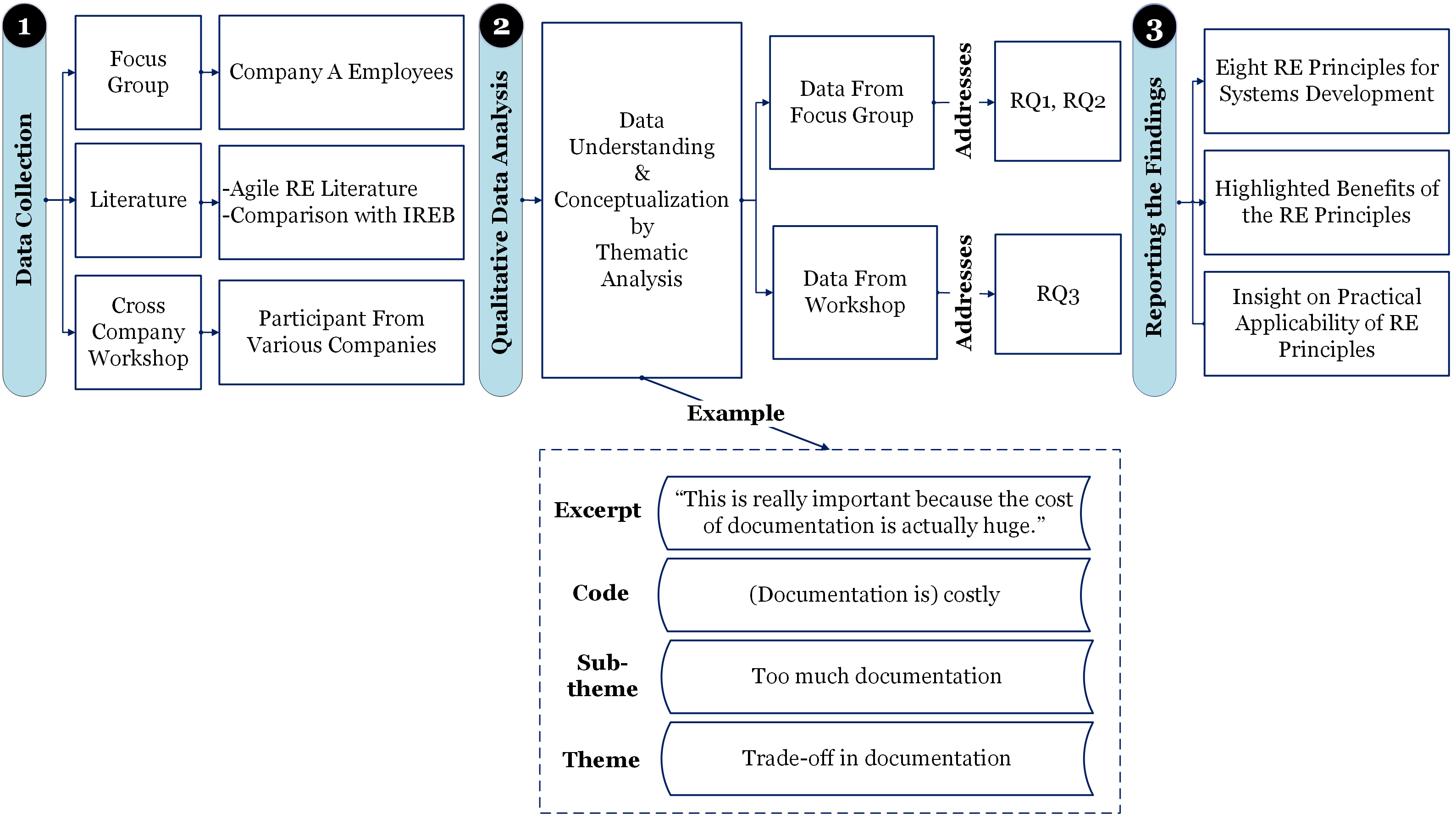}
\caption{Research method steps followed}
\label{fig:data_analysis}
\end{figure*}

We conducted thematic analysis following Braun and Clarke’s six-phase guidelines~\cite{braun2006}, supported by Nvivo and Microsoft Excel. To ensure methodological rigour, we incorporated inter coder agreement metrics and iterative team discussions. The process unfolded as follows: \begin{enumerate} \item \textbf{Familiarisation with the data:} All textual datasets, including interview transcripts, workshop notes, sprint reports, and synchronisation meeting logs, were reviewed multiple times by two researchers to achieve immersion and to flag potentially relevant statements. \textit{Example:} In an interview, one participant noted: “We often lose track of requirements when hardware and software teams don’t synchronise.” Both researchers independently marked this as relevant for traceability and alignment. \item \textbf{Generating initial codes:} Two researchers independently coded all textual data in Nvivo. Both inductive (data-driven) and deductive (informed by IREB principles and prior frameworks~\cite{IREB2024,Inayat2014,kasauli2017requirements}) approaches were applied. \textit{Example:} The above quote was coded as \texttt{requirements traceability problem}. Workshop data such as “developers are unclear about who owns which requirement” was coded as \texttt{unclear ownership}. \item \textbf{Inter-coder agreement:} To ensure coding reliability, 20\% of the dataset was coded by both researchers and compared. Cohen’s Kappa was calculated at $\kappa = 0.82$, indicating substantial agreement. Discrepancies were discussed and resolved through consensus, and the coding scheme was updated accordingly. The remaining data was then coded using the refined scheme, with regular calibration discussions to avoid coder drift. \item \textbf{Searching for categories:} Codes were iteratively clustered into conceptual categories in Excel to capture frequency and co-occurrence. \textit{Example:} Codes such as \texttt{requirements traceability problem}, \texttt{misalignment across teams}, and \texttt{fragmented documentation} were grouped into the category \texttt{alignment and traceability challenges}. \item \textbf{Constructing themes:} Categories were elevated into higher-level themes and sub-themes. \textit{Example:} The category \texttt{alignment and traceability challenges} was organised under the theme \texttt{Shared Understanding of Requirements}. Sub-themes included \texttt{traceability gaps} and \texttt{documentation inconsistencies}. \item \textbf{Reviewing and refining themes:} Themes were iteratively refined through collaborative research team discussions (involving two senior researchers and one early-career researcher) and validated against the research questions. Conflicts were resolved in joint interpretation sessions. \textit{Example:} Initially, \texttt{architectural dependencies} was grouped under \texttt{alignment}, but later refined into a standalone theme \texttt{Architectural Context} after repeated mentions in workshops that “system architecture decisions often drive requirement changes.” \item \textbf{Defining and naming themes:} Final themes were defined, named, and aligned with the derived RE principles. Each principle was linked back to the empirical evidence from the coded dataset, ensuring traceability between data and output. \textit{Example:} The theme \texttt{Shared Understanding of Requirements} was consolidated into the RE principle \textit{shared understanding}, supported by multiple quotes and categories documented in the replication package. \end{enumerate} \noindent Reliability was further enhanced through triangulation across multiple data sources (sprint reports, workshops, interviews, synchronisation meetings), inter-researcher coding, and member checking by company representatives. Figure~\ref{fig:data_analysis} illustrates how the data sources were mapped to each research question and how thematic analysis was performed. The \noindent \href{ https://doi.org/10.7910/DVN/RYHXJY}{\underline{\textcolor{blue}{\textit{replication package}}}}
 provides detailed thematic maps, coding hierarchies, and representative quotes.

\begin{figure}[ht]
\centering
\includegraphics[width=0.5\textwidth]{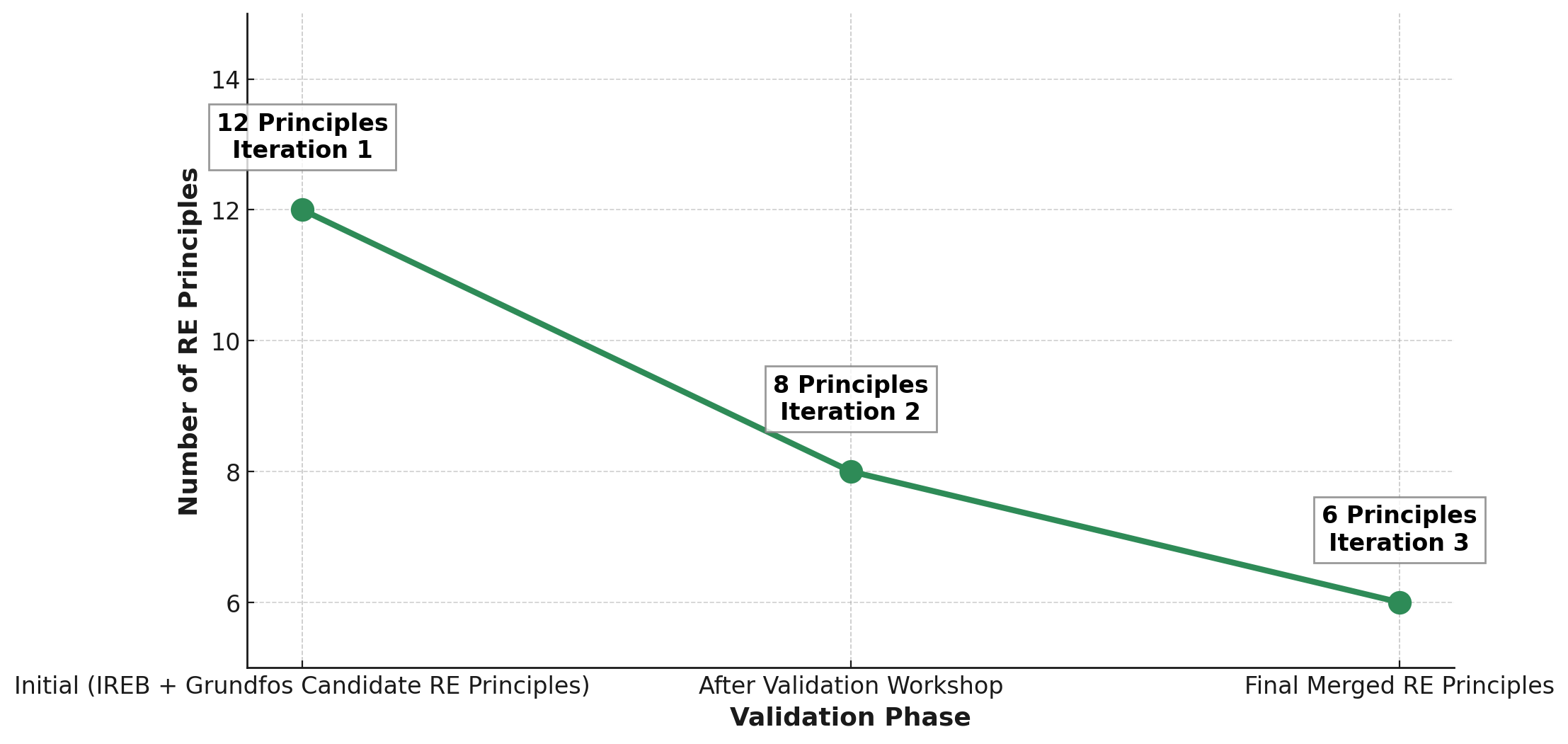} 
\caption{RE principles validations iterations.}
\label{fig:evolution_principles}
\end{figure}

\section{Findings }

Six \textit{RE principles} derived and implemented in \textit{Grundfos AB} were investigated to answer the \textit{RQ1}. \textbf{Figure}~\ref{fig:evolution_principles} illustrates the progressive refinement of RE principles throughout the study. As shown in \textbf{Tables} \ref{tab:combined_principles} and \ref{tab:workshop}, from 12 initial RE principles, overlaps, and redundancies were addressed in the cross-company workshop, reducing the set to 8 RE principles. For example, problem requirement solution was integrated into the broader principles of Validation and Customer needs alignment, while ownership-related practices were consolidated under Organisational roles and responsibilities.

In the final validation step, two further consolidations occurred: (1) Validation and Customer needs alignment were merged into a single principle of stakeholder-driven validation, and (2) Requirements evolution and Minimal documentation were combined into one principle balancing adaptability with lightweight artefacts. This refinement yielded six final RE principles, ensuring clarity, non-redundancy, and broad organisational applicability. The thematic maps used for the analysis can be found in 
\noindent\href{ https://doi.org/10.7910/DVN/RYHXJY}{\textcolor{blue}{\textit{\underline{Appendix(B)-rq1}}}}.

\subsection{(RQ1) Overarching RE principles in (Case Company)}

\textit{\textbf{(1) Know the architecture:}} Requirements should align with both system (functional, logical, physical) and organisational (roles, responsibilities) architectures. Without architectural context, requirements lack real-world applicability, impeding effective problem-solving. Knowledge of the system and its connections is essential for proper requirements management.

Consider a scenario where Grundfos AB is developing an IoT-enabled water pump system that interacts with subsystems like data analytics, cloud platforms, and mobile apps. To manage requirements, teams must align them with the system and organizational architecture. For instance, the pump sends data to cloud analytics, which alerts the mobile app. The requirements team must ensure real-time monitoring aligns with the system’s data processing, latency, and storage capabilities. Clear architecture and well-defined interface requirements enable teams to work independently yet stay aligned, reducing the need for frequent communication and speeding up decisions. As one participant noted, communication, though still necessary, becomes less frequent.

\begin{quote}
\small
\textit{“I believe [...] knowing the architecture and clear interface requirements help teams work aligned yet independently[...] reducing constant communication and increasing decision speed. Communication is still required, just less.”} —Systems engineer \& architect
\end{quote}

\textit{\textbf{(2) Democratize the RE job:}}
Requirements are often developed and updated by a few, limiting shared understanding. However, RE is a collaborative effort, not just an individual responsibility. Involving more people in RE work and delegating responsibilities fosters a shared understanding across the team. Consider a scenario where Grundfos AB needs to ensure accurate data processing and display on a mobile app for its water monitoring system. With complex hardware, analytics, and UI integration, it is essential for teams to align their work with requirements. Grundfos AB aims to democratize the RE process to improve collaboration, involving more team members in managing and updating requirements. For example, Grundfos AB holds bi-weekly alignment meetings with representatives from each area to review and update requirements. For instance, the hardware team may share sensor constraints, and the analytics team may discuss processing limits. These meetings ensure all teams understand the requirements and can adjust tasks, maintaining a shared development vision.

\textit{\textbf{(3) Aim at minimum viable documentation: }}Achieving the right level of requirements documentation is challenging. Too much documentation can waste resources, while too little risks knowledge loss. The ``minimum viable" amount varies based on team and task characteristics, product lifespan (longer lifespans often need more documentation), and the availability of experienced personnel (who can share knowledge directly). Despite these challenges, the goal should be to create useful and efficient minimum viable documentation. Grundfos AB focuses on essential documentation, covering only the system architecture overview, interface specifications, and critical safety requirements for the pump. This approach gives teams the necessary information for development and integration without overwhelming them with irrelevant details.

\textit{\textbf{(4) No perfect requirement exists: }} Good requirements have desirable features, as outlined in resources like the \textit{INCOSE Guide }\cite{incose2019guide}to writing requirements, and ideally, RE professionals should know these guidelines. However, pursuing ``perfect" requirements often leads to hesitation or excessive focus on minor details. Agile RE emphasises iterative revision, so requirements do not need to be perfect initially. Instead of aiming for perfection, people should focus on writing and continuously improving requirements throughout development. For example, as pump development progresses, the team revisits and refines requirements for each sprint, allowing flexibility and adaptation to new insights. 

\begin{quote}
\small
\textit{
``The term " perfection” includes [...] different requirements formats, e.g., one-liners/boilerplates [...] Systems Engineering vs. User Stories [...] as long as teams find them relevant/valuable. Tables and graphics are also excellent for creating a shared understanding.''— Systems engineer \& architect.}
\end{quote}

 The participant emphasizes the importance of rules and standards for high-quality requirements, noting that interpretations may vary among stakeholders.

\begin{quote}
\small\textit{
``It is excellent to know the INCOSE rules as guidance [... ]for what makes a good requirement. For instance, for quality criteria like ‘unambiguous’, one could ask, for whom? [...] the writer, the team, all teams in the project or company, suppliers, etc. It is difficult and often personal judgment [...], but the rules remain true and important guidance, whatever format is followed.''.— Systems engineer \& architect
}

\end{quote}

\textit{\textbf{(5) Refine when needed:}} Some in the organization confuse the stage-gate model with the RE process, thinking that requirements can only be updated in specific phases, whereas updates should occur as needed. Refinements in requirements and system details are encouraged, though frequent scope changes should be minimized to avoid confusion. 

While late refinement can be a risk in this way, it is equally risky not to change requirements later or not to decide that a requirement is good enough to proceed with development early enough.
\begin{quote}
\small\textit{
``[... Project teams are often paralyzed by the gates—decision points about risk handling and readiness. But risk is a gradual scale (not black/white)[...] there is no such thing as 'no risk."— Systems engineer \& architect}
\end{quote}

The participant notes varying ``risk" levels in requirements across domains: requirements tied to large production investments carry higher risk at concept freeze, while software features remain flexible for later adjustments. This highlights the varying risk levels at key project phases.

\textit{\textbf{(6) Know the ``why" and ``how":}} Requirements describe market needs and wants. Understanding why they should be managed ensures traceability from market needs to written requirements. Additionally, since requirements are closely tied to architecture, managing them effectively involves understanding this relationship and making requirements realistic and useful for problem-solving. 
Grundfos AB strives to translate market insights into specific, traceable requirements. For example, to meet energy efficiency needs, a requirement states that the pump should operate in an energy-saving mode during periods of low demand. For water quality, the pump should include sensors for pH, turbidity, and temperature, sending real-time data to a monitoring app. By mapping each market need to system requirements, Grundfos AB ensures traceability, showing how each feature meets customer demand and aligns with stakeholders' needs. This traceability ensures that requirements align with goals and expectations, as noted by a participant.
\begin{quote}
\small\textit{``[... ]'why' refers to Requirements Validation. Are we `building the right thing'? This validation is only possible with necessary traceability between requirements levels, enabling alignment with stakeholders.”— Systems engineer \& architect}
\end{quote}

The ``how" highlights the need for traceability between requirements and architecture to verify design accuracy and assess impacts, supporting iterative ``zig-zag" alignment.
\begin{quote}
\small\textit{``The `how' refers to the trace to architecture [...] crucial for impact assessment and design verification (did we build the thing right). This is only possible with req–architecture traceability [...] ties into the zig-zag pattern between requirements and architecture.” — Systems engineer \& architect}
\end{quote}

\subsection{(RQ2) The benefits of overarching \emph{RE principles} in (Case Company)}

This section presents three benefits of \textit{RE principles} for \textit{Grundfos AB}. The thematic analysis results for this RQ are shown in \noindent \href{ https://doi.org/10.7910/DVN/RYHXJY}{\textcolor{blue}{\textit{\underline{Appendix(C)-rq2}}}}.

\textit{\textbf{(1) A means of creating shared understanding: }}Growing system complexity challenges RE, requiring aligned requirements across various systems. Given the diverse backgrounds in RE, a shared language is essential for collaboration. RE principles can serve as this common language, enabling cross-disciplinary discussions.

\begin{quote}
\small
\textit{
``If you can get RE principles into play and [...]live in the organization, it provides a common language for something complex and hard to understand [...] creating a common language across digital, firmware, hardware, and mechanics [...] all relating to the same area with the same language, even if approached differently.” — System engineer \& Agile coach}

\end{quote}

\textit{\textbf{(2) Universal applicability: }}RE work varies across teams, each with unique processes, methods, and tools. While these specifics are hard to generalize, RE principles are tool-independent, providing high-level guidelines applicable across projects and teams.

\begin{quote}
\small
\textit{``We have many ways to improve; we can write a new process or enhance tools. But the principles are more tool-independent [...] not prescribing anything. [...] We made them universal to avoid black-or-white discussions." — Systems engineer \& architect}
\end{quote}

\textit{\textbf{(3)Help organization's improvement in RE: }}
Every organization faces unique RE issues and approaches. Establishing RE principles fosters discussions to identify key values within the organization, allowing teams to reflect on and improve their RE practices.

\begin{quote}
\small
\textit{``Principles are [...] a way to summarize our observations and guide colleagues toward a better understanding of our experience. [...] It’s another way to improve behavior." — Systems engineer \& architect}
\end{quote}

Moreover, RE principles help convey the implications of guidelines, enabling general lessons to inform specific RE tasks.

\begin{quote}
\small
\textit{``The INCOSE rules say a lot about good requirements writing. [...] You may believe the only way to write a requirement is as a one-liner, [...] but the rules also call for a coherent understanding[...] The principles help with that." — Software architect}
\end{quote}

%\begin{table}[h]
%\centering
%\caption{Mapping between key challenges and proposed RE principles}
%\label{tab:challenge-principle-mapping}
%\begin{tabular}{|p{5cm}|p{6cm}|}
%\hline
%\textbf{Challenge / Characteristic} & \textbf{Corresponding Principle(s)} \\
%\hline
%Cross-team coordination and alignment & Principle 1: Foster Alignment Across Teams \\
%\hline
%Architectural complexity and evolving system structures & Principle 2: Support Architectural Traceability \\
%\hline
%Distributed teams and stakeholder diversity & Principle 3: Facilitate Continuous Stakeholder Collaboration \\
%\hline
%Long product life cycles and evolving customer needs & Principle 4: Enable Continuous Requirements Refinement \\
%\hline
%Compliance, safety, and regulatory constraints & Principle 5: Ensure Compliance and Traceability \\
%\hline
%Balancing team autonomy and system-level consistency & Principle 6: Balance Autonomy and System-wide Consistency \\
%\hline
%\end{tabular}
%\end{table}

%Table~\ref{tab:challenge-principle-mapping} summarizes how key challenges/ characteristics of \hyperlink{para:important}{the large scale agile system development } inform the identified principles. 

\subsection{(RQ3) RE Principles: Insights from Literature and Practice}

\textit{\textbf{Insight from Literature Findings: }}The literature analysis with \cite{IREB2024} compared RE principles of Grundfos AB and IREB, highlighting both shared and unique priorities in agile systems development (see Table~\ref{tab:combined_principles}). 
Seven common principles, including understanding architecture, adapting to evolving requirements, reflecting customer needs, shared understanding, minimal documentation, and problem-solving focus, could serve as core RE principles. Grundfos AB emphasizes team-based RE roles, while IREB prioritizes systematic work, innovation, and rigorous validation, reflecting each organization’s distinct focus.
\begin{table*}[!ht]
\centering
\caption{Comparison between Grundfos AB and IREB RE Principles-Candidate Principles (prior to validation)}
\scriptsize % Reduces text size for compact presentation
\renewcommand{\arraystretch}{1.3} % Adjust row height for readability
\begin{tabular}{p{2cm} p{5.5cm} p{5.4cm} l} % Changed last column to left alignment
\toprule
\textbf{Category} & \textbf{RE Principle} & \textbf{Meaning of RE Principle} & \textbf{Derived from$^1$} \\
\midrule

\multirow{7}{*}{\textbf{Both(Grundfos\&IREB)}} 
& Systems architecture (context) & Understanding system architecture & A1, A6, I4 \\
& Evolution of requirements & Requirements change naturally & A4, A5, I7 \\
& Customers' needs & Reflecting customers' needs in RE & A6, I1 \\
& Shared understanding of problems \& solution & Building shared understanding of problems & A2, I3 \\ 
& Minimum viable documentation & Maintaining sufficient documentation & A3, I1 \\
& Problem, requirement, solution & Capturing problems through RE & A1, I5 \\
& Stakeholders' needs & Addressing stakeholders' needs in RE & A6, I2 \\

\midrule
\multirow{2}{*}{\textbf{ Grundfos AB}} 
& Delegating RM responsibilities & Involving multiple people in RE & A2 \\
& Organizational architecture & Defining RE roles within teams & A1, A2 \\

\midrule
\multirow{3}{*}{\textbf{ IREB}} 
& Validation & Emphasizing the validation of requirements & I6 \\
& Innovation & Going beyond stakeholders' requests in RE & I8 \\
& Systematic \& disciplined work & Valuing system \& discipline in RE processes & I9 \\

\bottomrule
\end{tabular}
\label{tab:combined_principles}
\end{table*}

\begin{figure}[ht]
\centering
\includegraphics[width=0.5\textwidth]{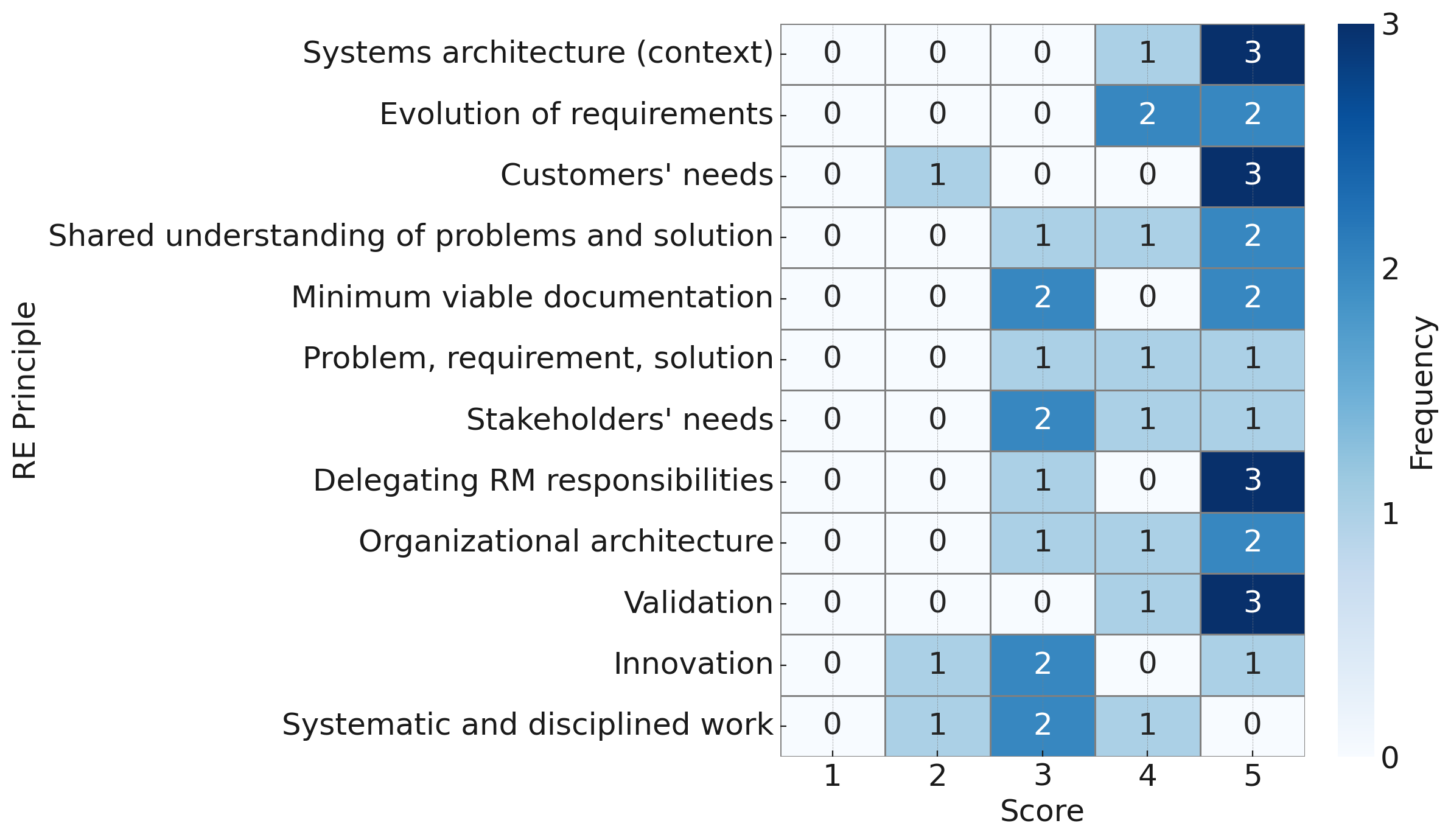} 
\caption{Distribution of response frequencies across RE Principles by score.}
\label{fig:heatmap}
\end{figure}

\textit{\textbf{Practitioners Insight from the Workshop: }} Workshop discussions resolved outstanding RE principle issues.  Table 3 summarizes each principle’s usefulness, while Figure~\ref{fig:heatmap} shows response distribution. The Mean indicates the average importance rating, while the Mode highlights the most common responses. Variance and standard deviation indicate the degree of variation in opinions, with lower values indicating greater agreement. The Rank summarizes the relative importance of each principle based on the mean, with ties noted in parentheses. 

Systems Architecture and Validation rated highest, followed by the Evolution of Requirements and Delegating RM Responsibilities for large projects. Customers’ Needs and Shared Understanding were valued but varied by the organization. Innovation and Systematic Work, unique to IREB, had lower ratings as participants viewed them as inherent to systems engineering. Full thematic analysis of the findings is  given in \noindent \href{ https://doi.org/10.7910/DVN/RYHXJY}{\textcolor{blue}{\textit{\underline{Appendix(D)-rq3.}}}}

\begin{table*}[!ht]
\centering
\caption{Workshop Results on the RE Principles}
\scriptsize % Makes the table text smaller for compact presentation
\renewcommand{\arraystretch}{1.2} % Adjust row height for readability
\begin{tabular}{p{2.8cm} p{4.5cm} p{0.9cm} p{0.9cm} p{1.2cm} p{1.0cm} p{0.9cm}} % Adjusted column widths
\toprule
\textbf{Category} & \textbf{RE Principle} & \textbf{Mean} & \textbf{Mode } & \textbf{Variance} & \textbf{Std Dev} & \textbf{Rank$^*$} \\
\midrule

\multirow{7}{*}{\textbf{Covered by Both }} 
& Systems architecture (context) & 4.75 & 5 & 0.25 & 0.50 & 1 (2) \\
& Evolution of requirements & 4.50 & 4, 5 & 0.33 & 0.58 & 2 (2) \\
& Customers' needs & 4.25 & 5 & 2.25 & 1.50 & 3 (3) \\
& Shared understanding of problems and solution & 4.25 & 5 & 0.92 & 0.96 & 3 (3) \\ 
& Minimum viable documentation & 4.00 & 3, 5 & 1.33 & 1.15 & 4 (2) \\
& Problem, requirement, solution & 4.00 & 3, 4, 5 & 1.00 & 1.00 & 4 (2) \\
& Stakeholders' needs & 3.75 & 3 & 0.92 & 0.96 & 5 (1) \\

\midrule
\multirow{2}{*}{\textbf{Grundfos AB}} 
& Delegating RM responsibilities & 4.50 & 5 & 1.00 & 1.00 & 2 (2) \\
& Organizational architecture & 4.25 & 5 & 0.92 & 0.96 & 3 (3) \\

\midrule
\multirow{3}{*}{\textbf{IREB}} 
& Validation & 4.75 & 5 & 0.25 & 0.50 & 1 (2) \\
& Innovation & 3.25 & 3 & 1.58 & 1.26 & 6 (1) \\
& Systematic and disciplined work & 3.00 & 3 & 0.67 & 0.82 & 7 (1) \\

\bottomrule
\end{tabular}
\label{tab:workshop}
\\[0.1cm] 
\footnotesize{$^*$ The means are used to calculate the ranks. The number of ties, including itself, is presented in parentheses.}
\end{table*}

\textit{\textbf{(Rank 1) Systems Architecture (Context), Validation: }}
 Both principles ranked at the highest, with a small variance in the responses, demonstrating their importance in systems engineering across diverse domains. Especially the importance of \textit{Validation} was emphasized multiple times. 
 
\textbf{\textit{Key-takeaway(arch. context)}:}
Effective large-scale development requires balancing team autonomy with architectural coherence. Structured governance, shared models, and regular inter-team alignment support consistent decision-making and seamless subsystem integration.

\begin{quote}
\small
\textit{``As a system engineer with RE, you want to be as close to validation as possible[...] you have a strong development environment[...] you know that the customer will be happy in advance; so there are fewer surprises." —P1 }

\end{quote}
Interestingly, P4, a founding member of the RE principles at Grundfos AB, identified \textit{Validation} as crucial to systems engineering, despite it not being included in the original principles.

\textbf{\textit{Key-takeaway (validation):}} In large-scale systems, validation is critical for early and continuous verification. It helps detect issues early, reducing rework and failures. Practices like integration sprints, automated testing, and shared test environments improve quality and coordination throughout development.

\textit{\textbf{(Rank 2) Evolution of Requirements, Delegating RE responsibilities: }}
As a core Agile value, \textit{Evolution of Requirements} is natural in agile development.

\textbf{\textit{Key-takeaway (evolution of requirements):}} In large-scale system development, evolving requirements are essential to address changing needs and insights. Coordinated updates, stakeholder feedback, and traceability across design, tests, and regulations ensure consistent and compliant adaptation.

\textit{Delegating RM responsibilities} is essential in large projects, where RE can not be a single person's role, and its value varies by requirement level, which may explain the high variance in responses.

\begin{quote}
\small
\textit{``We have customer requirements, system-level requirements, and lower-level requirements[...]some are delegated, others more streamlined." — P3 }
\end{quote}
\textbf{\textit{Key-takeaway (delegating RE responsibilities):}} In large-scale agile development, delegating requirements management to roles like product owners improves scalability and responsiveness. To ensure coherence, this should be supported by clear roles, shared tools, synchronization, and system-level oversight.

\textit{\textbf{(Rank 3) Customers' needs, Shared Understanding of Problems and Solutions, and Organizational Architecture:}}
For \textit{Customer's needs}, workshop discussions showed that varying interpretations of ``customers" might explain the high response variance.  

\textbf{\textit{Key-takeaway (customers' needs):}} In large-scale agile development, meeting evolving customer needs requires continuous stakeholder engagement and structured feedback. Roles like business owners, regular reviews, and MVP-based validation help manage complexity and ensure a shared understanding across distributed teams.

 \textbf{\textit{Key-takeaway (shared understanding of problems and solutions):}} Shared understanding among teams and stakeholders is crucial for consistent requirement interpretation and customer alignment. Early involvement of both technical and business stakeholders—through joint workshops, collaborative modeling, and shared documentation—helps prevent costly misunderstandings by fostering a common mental model and unified system vision.

Notably, \textit{Organizational Architecture}, unique to Grundfos AB, was also valued across other company contexts.

\textbf{\textit{Key-takeaway (organizational architecture):}} A clear organizational structure is vital for effective RE in large-scale agile development. Defining roles, communication channels, and decision-making processes, along with practices like cross-team coordination and shared repositories, helps manage complexity and maintain alignment across teams.

\textit{\textbf{(Rank 4) Aligning with Agile values:}} \textit{Minimum Viable Documentation }was emphasized by both Grundfos AB and IREB and approved by practitioners. While \textit{Problem}, \textit{requirement}, and \textit{solution }are generally valued, a few participants questioned their suitability as an RE principle.

   \begin{quote}
\small
    \textit{``I don’t think it is a good principle as it is a native part of practice if you are committed to Systems Engineering."  — P4}
   
    \textit{``While (the principle is) true, we manage without explicitly referring to this principle, so I'm not sure how useful it is."  — P2}
\end{quote}

            \textbf{\textit{Key-takeaway (aligning with agile values):}} Aligning RE with agile values requires balancing flexibility and structure, achieved through lightweight documentation, ongoing stakeholder involvement, and integrating RE into agile practices like backlog refinement and sprint reviews to support iterative requirement evolution.

\textit{\textbf{(Rank 5) Stakeholders' needs: }} While both \textit{Customers' needs} and \textit{Stakeholders' needs} are important in RE, participants valued the former more. Discussions indicated that customers are often seen as more critical or are considered identical to stakeholders, explaining the difference in means.

\textbf{\textit{Key-takeaway (stakeholders' needs):}} In large-scale system development, engaging diverse stakeholders through structured analysis and regular interaction is essential to capture comprehensive requirements, ensuring both functional and non-functional needs are effectively addressed.

\textit{\textbf{(Rank 6 \& Rank 7) Innovation \& Systematic and disciplined work: }}
Similar to `\textit{Problem, requirement, solution}', most participants did not see a clear benefit of having these two as RE principles. A participant stated that both of them automatically follow systems engineering work. As for \textit{Innovation}, a participant expressed that it is a requirement for the business aspect rather than for systems engineering. Still, the participants agreed on their importance in systems engineering, and one significantly favored the principle, accounting for the high variance. P1 stated about \textit{Innovation} as follows:

\begin{quote}
\small
    \textit{``I think it's not necessarily a principle for agile development[...], but it goes hand in hand with it[agile]—the team has innovation. The opposite is centralized control, where engineers do exactly what the manager tells them[...]such development works, but not as efficiently in my experience."  — P1}

\end{quote}

\textbf{\textit{Key-takeaway (innovation \& systematic and disciplined work):}} Enabling innovation in large-scale agile development involves balancing exploration with system stability and compliance. RE should support this by allowing experimentation through practices like innovation sprints and hackathons, while governance structures ensure alignment with architecture and value. This fosters continuous learning without compromising system coherence.

In addition to the principles in Table~\ref{tab:workshop}, several participants suggested \textit{`Clear responsibilities of requirements'} as an RE principle, highlighting a stronger focus on practical implementation within \textit{Organizational architecture}.

\begin{quote}
\small
    \textit{``From my experience in agile[...] sometimes things are less clear or up to the team. However, if no one takes responsibility, it becomes tricky to achieve a shared understanding or effective communication around requirements. So, I think it helps to have some agreed-upon responsibilities and requirements." — P3}
\end{quote}

Workshop participants agreed on an ideal range of five to nine \textit{RE principles} for their organizations. 
The needs of a specific company may differ. Thus, we recommend starting from the following \textit{six principles} for broad applicability in large-scale agile systems engineering: 
\textit{Systems Architecture (Context), Validation, Evolution of Requirements, Clearly Define \& Delegate RM Responsibilities, Shared Understanding of Problems and Solutions, and Minimum Viable Documentation}.

\section{  Discussion and Conclusions   }

This longitudinal industry study report identifies Six RE principles that enhance requirements management in large-scale agile development at Grundfos AB, based on triangulated insights from interviews, workshops, and literature.

While several of the identified RE principles reinforce existing knowledge, others extend beyond prior frameworks. For example, principles such as \textit{validation}, \textit{requirements evolution}, and \textit{customer needs alignment} resonate strongly with both the IREB Handbook~\cite{IREB2024} and Inayat et al.’s ~\cite{Inayat2014} systematic review of agile RE practices, which emphasise stakeholder involvement, iterative refinement, and prioritisation. Similarly, challenges related to \textit{roles and responsibilities} and \textit{shared understanding} are well-documented in Kasauli et al.’s~\cite{kasauli2017requirements} taxonomy of 24 RE challenges in large-scale agile system development, underscoring their relevance in large-scale agile settings.  However, two principles \textit{architectural context} and \textit{organisational roles and responsibilities} go beyond the coverage of IREB, Kasauli, and Inayat. These principles highlight the necessity of explicitly managing system-level architectural dependencies and clarifying multi-level organisational accountability, both of which are underrepresented in prior frameworks but emerged as critical in our longitudinal case. Furthermore, our emphasis on \textit{minimal documentation} addresses the tension between agile practices that advocate lightweight artefacts ~\cite{Inayat2014} and the regulatory and traceability demands in large-scale system development, offering a more balanced principle than is reflected in existing frameworks. 

Our principles provide not only confirmation of established insights but also novel contributions that target gaps in managing architecture, organisational accountability, and documentation trade-offs in large-scale agile environments.

The RE principles at \textit{Grundfos AB}, such as “\textit{Know the Architecture},” “\textit{Democratise the RM Job},” and “\textit{Minimum Viable Documentation},” align with core agile values collaboration, iteration, and customer focus—as highlighted in the Agile Manifesto~\cite{agileManifesto}. These principles help manage the complexity of large-scale projects by promoting shared understanding and a lightweight documentation strategy.  Unlike traditional RE frameworks that stress comprehensive documentation and upfront requirements~\cite{incose2019guide}, \textit{Grundfos AB} adopts a “just enough” approach, reducing overhead while preserving critical knowledge. This reflects the concerns raised by Inayat et al.~\cite{Inayat2014}, who argued that excessive documentation can limit agility in large-scale projects. The identified principles also partially align with the IREB RE principles~\cite{IREB2024}, particularly regarding stakeholder alignment and architectural awareness. However, while IREB's principles are context-independent and systematic, they are not specifically tailored for large-scale agile coordination. This study complements IREB’s guidelines by demonstrating how such principles can be interpreted and adapted in agile environments.

A notable principle—“\textit{Delegating RM Responsibilities}”—emphasizes distributed ownership of requirements, aligning with findings from Kasauli et al.\cite{kasauli2017requirements}, who highlight its role in maintaining alignment across agile teams. This collaborative model also resonates with \cite{Inayat2014}, who underscore the importance of coordination in multi-team settings. Finally, compared to IREB’s structured principles like “\textit{Systematic and Disciplined Work}” and “\textit{Innovation Beyond Stakeholder Requests}”\cite{IREB2024}, \textit{Grundfos AB} emphasizes adaptability, as seen in its “\textit{Refine When Needed}” principle. This reflects Agile’s iterative nature and contrasts with traditional sequential RE processes, supporting Saeeda et al. ~\cite{saeeda2023challenges}' observations on the need for flexible RE in dynamic settings.
From our longitudinal study, we learned that high-level principles provide more sustainable guidance than rigid processes in large-scale agile settings. Embedding architectural context, delegating RE responsibilities, and aiming for minimum viable documentation were crucial for alignment and scalability. Continuous validation and maintaining traceability to both “why” (customer value) and “how” (architecture) emerged as critical enablers of effective RE practice.

\textbf{Validity threats:} This study recognises several potential threats to validity ~\cite{Runeson2009}.

\textbf{Construct Validity:} While participants may have emphasised their own organisations, context-specific bias was mitigated through triangulation across interviews, literature, and cross-company workshops. In longitudinal case studies such as ours, the focus is on depth and coverage rather than participant count. The participants, as senior leaders coordinating 50+ developers each, were continuously engaged in synchronisation meetings and directly involved in implementing the RE principles. This ensured validation was grounded in lived organisational practice rather than abstract opinion. Following established case study guidance~\cite{Runeson2009}, information-rich participants and contextual depth were prioritised over sample size. External validity was further strengthened by cross-company validation with Bosch, Ericsson, and Volvo Cars, confirming that the principles generalise beyond the Grundfos AB context.

\textbf{Internal Validity:}
 We followed established thematic analysis guidelines~\cite{braun2006, nowell2017}, with two researchers independently coding the data and resolving discrepancies through iterative discussions. Second, triangulation was achieved across multiple data sources, including more than 25 sprints, approximately 320 synchronisation meetings, seven cross-company workshops, and final retrospective focus groups, reducing reliance on any single perspective. Third, member checking was performed by sharing synthesised interpretations with participants throughout the project to ensure accuracy and credibility of the findings.  Anonymous Mentimeter ratings reduced bias and groupthink during principle prioritisation, while continuous team discussions avoided single-researcher dominance. The principles were progressively implemented and refined over a five-year period, providing longitudinal validation in practice. Together, these measures strengthen internal validity by grounding results in systematic analysis and sustained organisational evidence

\textbf{External Validity:}  
This study offers actionable insights for both researchers and practitioners. The RE principles were derived in close collaboration with \href{https://www.grundfos.com/}{Grundfos AB}, a multinational organisation employing more than 100 agile teams across hardware, software, and system integration domains. While grounded in this case, external validity was strengthened through cross-company validation with experts from Bosch, Ericsson, and Volvo Cars. These organisations operate in diverse large-scale, safety-critical, and multidisciplinary domains, yet confirmed the relevance and transferability of the proposed principles.  

Moreover, the principles were validated not as abstract concepts but as practices already implemented across more than 100 teams during the five-year study, which provides stronger evidence of applicability than post-hoc evaluations. The challenges addressed—coordination, alignment, and complexity are consistent with those reported in prior large-scale agile research~\cite{r4dikert2016challenges,r9edison2021comparing}, suggesting that the principles target issues common across many industrial contexts. While tailoring is necessary for each organisational environment, the findings offer generalisable guidance for other large-scale agile settings facing similar demands for traceability, architectural alignment, and cross-team coordination.

\textbf{This study offers actionable insights for both researchers and practitioners.} First, a shared understanding of the problems was found to be essential for coordination. Misaligned mental models across teams often lead to inconsistent requirements. Techniques such as collaborative modelling, joint elicitation workshops, and backlog refinement help establish early consensus and reduce integration risks~\cite{Dingsoyr2014}. Second, aligning with agile values requires striking a balance between flexibility and structure. The use of minimum viable documentation, as recommended by participants, supports agility without sacrificing traceability~\cite{ r4dikert2016challenges}. Third, principles like \textit{Systems Architecture (Context)} and \textit{Validation} were rated as critical. These support complexity management and coherence through governance mechanisms such as architecture boards, shared test environments, and validation sprints. Fourth, \textit{Evolution of Requirements} is vital in dynamic settings. Continuous refinement and stakeholder feedback loops allow adaptation while preserving architectural and compliance integrity~\cite{Inayat2014}. The principle \textit{Delegating RM Responsibilities} promotes scalable RE by encouraging distributed ownership through roles such as product owners and feature leads~\cite{Dingsoyr2014, treqs}, supporting autonomy while maintaining alignment. Customer needs were prioritized slightly higher than stakeholder needs, reflecting agile's user-centric focus~\cite{r10beck2001manifesto}, yet recognizing diverse stakeholder requirements remains essential~\cite{kasauli2017requirements}.

\textit{Organisational Architecture} also emerged as an enabler for effective RE. Structuring around architectural components or value streams, with cross-team synchronization roles, supports coordination without undermining agility~\cite{r4dikert2016challenges}. The principle of \textit{Innovation} highlights the need for exploration alongside stability. Practices like innovation sprints and exploratory spikes embed innovation into large-scale agile workflows~\cite{orosz2023scaling}. Finally, addressing \textit{Requirements Synchronization} through visual models, shared artifacts, and multi-team backlog refinement supports cross-team alignment, responding to challenges identified in prior research~\cite{conboy2019implementing,r9edison2021comparing}. 

\begin{acks}
    
 This research was supported by the Software Center (Project 27) and Vinnova FFI (Project FAMER, 2023-00771). 
\end{acks}

%\subsection{Acknowledgments}
%This research was supported by the Software Center (Project 27) and Vinnova FFI (Project FAMER, 2023-00771). 

\bibliographystyle{ACM-Reference-Format}
\bibliography{sample-base}

\end{document}